\documentclass{article}
    \usepackage[english]{babel}
    \usepackage[margin=1in]{geometry}
    \usepackage{multicol}
    \usepackage{background}
    \usepackage{hyperref}
    \usepackage{graphicx}
    \usepackage[space]{grffile}
    \usepackage{amsmath} 
    \usepackage[labelfont=bf]{caption}
    
    \SetBgScale{4}
    \SetBgColor{gray}
    \SetBgAngle{90}
    \SetBgContents{arXiv: some other text goes here}
    \SetBgPosition{-1.5,-10}
    
\begin{document}
\normalfont (Journal of the Geological Society of India, In Press)
    \bigskip
    \bigskip
    \bigskip
    
    {\centering

    {\bfseries\Large Analysis of Pre- and Post-Monsoon Suspended Sediments in the Gulf of Kachchh, India Using Remote Sensing\bigskip}

    Mukesh Gupta \\
    (ORCID iD: \url{http://orcid.org/0000-0002-8955-6931}) \\
    \bigskip
      {\itshape
    Earth Sciences and Hydrology Division, Marine and Water Resources Group, Space Applications Centre, Indian Space Research Organisation, Ahmedabad-380 015, Gujarat, India \\
    Presently at: Meteorological Research Division, Data Assimilation and Satellite Meteorology Research Section, Science and Technology Branch, Environment Canada, Dorval H9P 1J3, Quebec, Canada
\\
     \bigskip
     Corresponding author e-mail address: guptm@yahoo.com
    
    \normalfont (Received: 4 May 2013; Revised form accepted: 26 August 2013)
   
      } \bigskip
    }

    \begin{ }
    \small A comprehensive study of satellite-derived suspended sediment concentration (SSC) during pre- and post-monsoon has been conducted with full-month cycles of tidal responses to study the suspended sediment dynamics in the Gulf of Kachchh. Tidal data were interpreted in conjunction with the OCEANSAT-1 ocean color monitor (OCM)-derived SSC for pre- and post-monsoon. The analysis of the data shows that the Gulf is predominantly affected by the tidal changes. The average SSC during pre-monsoon were 30.8 mg/l (high tide) and 24.1 mg/l (low tide); and during post-monsoon 19.7 mg/l (high tide) and 21.8 mg/l (low tide). The only little monsoonal influence is seen when Indus River discharges sediments during pre-monsoon due to increased sediment flux from its origin, Himalayas in spring (February\textendash April) as compared to less sediment discharge observed during winter (November\textendash December). The pre-monsoon SSC images show overall high suspended sediments whereas post-monsoon SSC images show comparatively low SSC. The use of enhanced resolution ocean color satellite data (\textless360 m spatial resolution) for deriving higher SSC (\textgreater40 mg/l) for moderate/normal/high monsoon years under similar tidal conditions, and for quantifying sediment dispersal and dynamics and its validation is suggested as a future avenue of research.
    \bigskip


    \end{ }

\begin{ }
\section*{\normalsize INTRODUCTION}
Sediment transport is a process by which sediments that are in suspension or at the bottom get transported to other geographic locations under the influence of physical forces such as currents, waves, wind, sea surface temperature etc. Suspended sediments have a direct impact on the underwater light availability for primary production, coastal erosion and accretion, transport of minerals and pollutants to the oceans. These have been studied by coastal engineers for the construction of coastal structures like jetties, breakwaters etc. The usage of satellite remote sensing data for the dispersal and distribution of suspended sediments has proven to be useful because of high temporal images (every alternate day) at spatial resolution of 360 m available from ocean color monitor (OCM) onboard OCEANSAT-1 \cite{pradhanetal2004}.\\
	\indent Gulf of Kachchh (also spelled as Kutch, Cutch, or Kachh) is a seismically active, funnel-shaped, macro-tidal (mixed semi-diurnal) environment with tidal height variations of 3\textendash 7 m \cite{chauhanandvora1990}\cite{shetye1999}. Singh et al. (2001)\cite{singhetal2001} reported the effect of Gujarat earthquake (magnitude 7.8) on the suspended sediment concentration (SSC) in the coastal waters of Gujarat. No major earthquake or tectonic movement has been reported since 2001 in the region that could potentially vertically mix the coastal waters. The coastal region of the Gulf encompasses a variety of landforms such as tidal flats, sand spits, saltpans, river inlets, and alluvial plains \cite{patelanddesai2009}. The major source of sediments and minerals in the Gulf is Indus River (about 100 km further Northwest), which brings nearly 435 million tons of suspended sediments load every year \cite{kunteetal2003} \cite{chauhanetal2006} \cite{trivedietal2012}. The Gulf is highly turbid with SSC during October\textendash November ranging between 0.5\textendash 674 mg/l \cite{ramaswamyetal2007}. The Gulf experiences strong longshore currents at its mouth discernible through sediment dispersal patterns observed by OCEANSAT-1 \cite{pradhanetal2004}. The sediment dispersal patterns are seasonal and are controlled by the current velocity distribution. The tidal circulation time is about 10 hours in the Gulf \cite{unnikrishnanandluick2003}. Also, the strength of tidal currents in the Gulf of Kachchh is further enhanced due to its channel geometry \cite{shetye1999}.\\
	\indent Satellite-based monitoring of suspended sediments is of significance due to its greater temporal and spatial coverage as compared to sea- or ground-truth observations. Ocean color satellite sensors provide a wealth of information in different spectral channels about the various constituents of seawater. Ocean color imagery-derived products of SSC can be further used for estimating dispersal, extent, and quantification of suspended sediments in the Gulf of Kachchh. OCEANSAT-1 OCM data have been demonstrated for its widespread utility in sediment dispersal dynamics studies \cite{pradhanetal2004}, underwater bed forms, and sediment plume \cite{kunteetal2003}. OCM-derived SSC products have also been used to simulate and predict SSC in the Gulf of Kachchh \cite{ramakrishnanandrajawat2012}. Chauhan et al. (2007) \cite{chauhanetal2007} used the high spatial resolution of OCM data to find sources, sinks, and dispersal pathways of suspended sediments in the Gulf of Kachchh. SSC is retrieved from satellite-detected water-leaving radiance in the electromagnetic channels that are sensitive to reflection/absorption from suspended sediments. Tassan (1994)\cite{tassan1994} derived an empirical relationship to estimate SSC (up to 4.6 mg/l) based on radiometric observations of the reflectance of suspended sediments. The algorithm has been modified for deriving SSC up to 40.0 mg/l using satellite remote sensing data \cite{pradhanetal2004}. Other algorithms are also available to retrieve SSC up to 4.6 mg/l from SeaWiFS data \cite{kunteetal2005}. The SSC (up to 200 mg/l) retrieval algorithm using inherent optical properties such as diffuse attenuation coefficient was also proposed \cite{pradhanetal2003}.\\
\begin{table}[h!]
\begin{center}
\caption{Specifications of OCEANSAT-1 ocean color monitor (OCM) sensor.} 
\label{tab:table1}

\begin{tabular}{ l r} 
\hline
System Parameter & Value \\ \hline
Spectral range & 404\textendash 882 nm \\
No. of channels	 & 8 \\
Wavelengths (SNR)	& Channel 1: 404\textendash 423 (340.5) \\
 & Channel 2: 431\textendash 451 (440.7) \\
 & Channel 3: 475\textendash 495 (427.6) \\
 & Channel 4: 501\textendash 520 (408.8) \\
 & Channel 5: 547\textendash 565 (412.2) \\
 & Channel 6: 660\textendash 677 (345.6) \\
 & Channel 7: 749\textendash 787 (393.7) \\
 & Channel 8: 847\textendash 882 (253.6) \\
Satellite altitude & 720 km \\
Spatial resolution & 360 m × 236 m \\
Swath & 1420 km \\
Repeat cycle & 2 days \\
Quantization & 12 bits \\
Equatorial crossing time & 12 noon \\
Along track steering (to avoid Sun glint) & 20° \\
MTF (at Nyquist) & \textgreater 0.2 \\
Transmission frequency & X-band \\
Data rate & 20.8 kbps \\
Launch date & 26 May 1999 \\
End of mission & 08 August 2010 \\ \hline 
\footnotesize SNR: Signal-to-Noise Ratio; MTF: Modulation Transfer Function
\end{tabular}
\end{center}
\end{table}

	 \indent OCEANSAT-1 (also known as IRS-P4) is the first Indian satellite primarily built for ocean applications. It weighed 1050 kg and was placed in a polar Sun-synchronous orbit at an altitude of 720 km, launched by PSLV-C2 rocket from Satish Dhawan Space Centre (SDSC), Sriharikota, India on 26 May 1999. The satellite carried OCM sensor among others for oceanographic studies. OCM acquired imageries in eight different spectral channels (404\textendash 882 nm) every alternate day (Table~\ref{tab:table1}). The mission ended on 08 August 2010 after serving for 11 years and 2 months.\\
	\indent Indian Meteorological Department (IMD), Government of India declared the monsoon in 2004 deficient \cite{rajeevanetal2007}. The absence of normal monsoon in 2004 gave a unique opportunity to study the role of external/other factors responsible for suspended sediments (e.g. Indus River sediments brought from the Himalayas, tidal forces) into the Gulf during the months otherwise identified as pre- and post-monsoon. Parallel to the main objectives given below, we make use of the scarcity of monsoon of 2004 to find through satellite data, the sources of suspended sediments in the Gulf. In this context, it is the aim of the paper to find out the behavior of SSC values before and after the monsoon considering full tidal cycles. The main objectives of this study are identified as follows:\\
\indent1. How does SSC change in the Gulf of Kachchh during pre- and post-monsoon for full tidal cycles based on remote sensing estimates?\\
\indent2. To exploit satellite remote sensing for monitoring the extent, dispersal, and dynamics of suspended sediments in the Gulf.\\

\begin{figure}[htbp]
\centering \includegraphics[height=11cm]{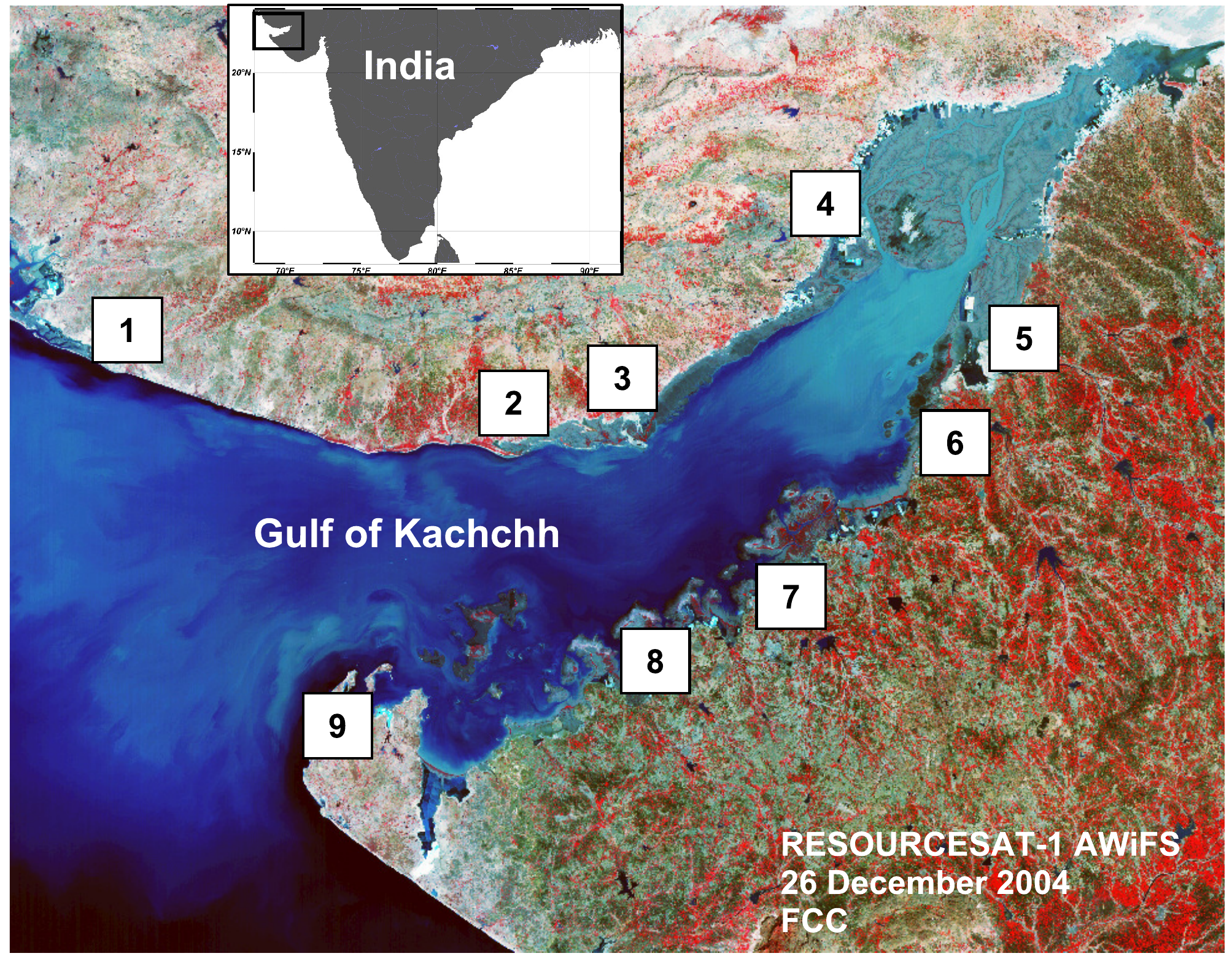} 
\caption{\small RESOURCESAT-1 AWiFS standard false color composite (FCC) (B2: blue, B3: green, B4: red) image of Gulf of Kachchh acquired on 26 December 2004. Various port locations (for tide measurements) shown in solid squares are: (1) Jakhau, (2) Mandvi, (3) Mundra, (4) Kandla, (5) Navlakhi, (6) Jodiya, (7) Sikka, (8) Kalubhar, and (9) Okha.} \label{fig:figure1}
\end{figure}

\section*{\normalsize STUDY AREA}
	The study area lies in the West coast of Indian state of Gujarat, in the northeastern stretch of Arabian Sea, between $68^{\circ}20^{\prime}\textendash 70^{\circ}40^{\prime}$ E and $22^{\circ}15^{\prime}\textendash 23^{\circ}40^{\prime}$ N covering an area of about 7300 km$^2$ (Fig.~\ref{fig:figure1}) \cite{rajawatetal2005}\cite{rajawatetal2003}. It sprawls about 180 km in length in East\textendash West direction, and about 70 km at the mouth to 1--2 km at the creeks in the East \cite{ramaswamyetal2007}. The depth varies from 60 m near the mouth to 20 m at the eastern end of the Gulf. The Gulf of Kachchh has major ports of Kandla, Mandvi, Mundra, Navlakhi, and Okha. There are no rivers, tributaries, or any fresh water source to the Gulf except the Indus River, which lies about 100 km Northwest from the mouth of the Gulf. \\

\section*{\normalsize METHODS}
\subsection*{\normalsize Data Used}
	A total of 41 OCEANSAT-1 sequential OCM images were used for pre- (February, March, and April 2004) and post-monsoon (November and December 2004) analysis of sediment dynamics in the Gulf of Kachchh depending on low and high tide conditions (Table~\ref{tab:table2}). National Remote Sensing Centre (NRSC), Hyderabad, India provided the satellite imageries. The imageries were provided as raw radiance image files, which were geo-corrected and were readied for SSC retrieval after atmospheric correction in Geographic Lat/Long coordinate system and Modified Everest datum. Six images (8, 14 February, 9, 11 March, and 8, 24 April) in the pre-monsoon and four images (12, 20 November, and 12, 18 December) in the post-monsoon were used to compute SSC in response to full tidal cycle. These 10 cloud-free images were selected in order to match the high or low tide conditions during a month for a given satellite pass. Remaining 31 SSC images were used for studying the suspended sediment dispersal patterns as observed through sequential satellite images \cite{gupta2015}.

\subsection*{\normalsize Atmospheric Correction}
	The raw imageries before preparing SSC products were corrected for atmospheric variability. It is well known that in oceanic remote sensing, the total signal received at the satellite sensor is dominated by radiance contribution through atmospheric scattering processes and only 8--10\% signal corresponds to oceanic reflectance \cite{chauhanetal2003}. It is known that for near infrared channels the water-leaving radiance is approximately equal to zero. The top of the atmosphere radiance in OCM wavelengths, 765 and 865 nm, mainly corresponds to contribution coming only from the atmosphere, therefore, the water-leaving radiance in 765 and 865 nm channels can be safely assumed to be zero. A relationship is obtained for the spectral behavior of the aerosol optical depth using wavelengths 765 and 865 nm in OCM data \cite{chauhanetal2003}. An exponential relationship for spectral behavior of aerosol optical depth has been used in the atmospheric correction algorithm. The aerosol optical thickness is extrapolated to visible channels using this exponential relationship. A detailed description of atmospheric correction applied to OCM scenes and further references are provided elsewhere \cite{chauhanetal2003}.

\subsection*{\normalsize Retrieval of SSC}
	The satellite images were analyzed using ERDAS Imagine 8.5 software. The imageries were geometrically corrected using the satellite ephemeris data. Tassan$'$s (1994) \cite{tassan1994} algorithm was modified to estimate the SSC up to 40 mg/l from the satellite radiances \cite{pradhanetal2004}. The SeaWiFS (Sea-viewing Wide Field-of-view Sensor) Data Analyses Software (SeaDAS 4.0) was used to color-code the SSC products. The SSC products were derived using water-leaving radiances at wavelengths 490, 555, and 670 nm. The radiance images were converted into remote sensing reflectance images using the method given by Chauhan et al. (2003) \cite{chauhanetal2003}. The algorithm used for retrieval of SSC is given in Equations (1) and (2) \cite{pradhanetal2004},
\begin{align}
	\log S=1.83+1.26\log X_s,      \; 0.0\le S\le 40.0 \\
	X_s=[Rrs(555)+Rrs(670)]\times \left[ \frac{Rrs(555)}{Rrs(490)}\right]^{0.5}.
\end{align}
where $Rrs(\lambda)$ is remote sensing reflectance at respective wavelengths ($\lambda$), $S$ is the SSC in mg/l. The retrieval accuracy of SSC using OCM data is within 15\%. The contribution to spectral reflectance comes from the volume of the water column. Water-leaving radiance as detected by the satellite sensor has contributions from the water constituents (e.g. chlorophyll, colored dissolved organic matter, and mineral sediments), particle size distribution, and light scattering/absorption from each constituent \cite{nanuandrobertson1993}. The depth to which remote sensors can detect SSC is modified by above constituents and available light. A typical range of this depth could be from a few centimeters (highly turbid: Case 2 waters) to several meters (open ocean: Case 1 waters) \cite{forgetetal2001}\cite{ouillon2003}. Ouillon (2003) \cite{ouillon2003} has reported the total suspended matter concentration of 6 mg/l at a depth of about 3 m. In the present study, it is safe to assume that the reflectance from suspended sediments is predominantly from the surface as the Gulf is highly turbid with SSC above 40.0 mg/l with little light penetration into the water column.

\subsection*{\normalsize Tide Data Analysis}
	Tide data were taken from Indian Tide Table (2004) prepared by Survey of India.  Tide graphs were plotted and the corresponding OCM images were analyzed and interpreted. Since OCM pass is at 12:00 noon IST (Indian Standard Time), the tide conditions closest to 12:00 noon were identified and the tidal heights were selected. The first highest and the first lowest values of tidal heights were picked. In case the date of occurrence of high/low tide did not match with the OCM date of pass, the subsequent highest and subsequent lowest values of tidal heights were selected to match with the OCM date occurrence \cite{gupta2015}. The high and low tides in pre-monsoon were 3.7 m (6 April) and 1.1 m (19 March), respectively. Similar high and low tides in post-monsoon were observed to be 3.4 m (12 November) and 1.2 m (20 November), respectively.
\bigskip
\begin{table}[h!]
\begin{center}
\caption{OCEANSAT-1 OCM image (total 41) dates (2004) used for suspended sediment concentration retrieval. Bolded dates (total 10) were used for SSC and sediment extent computations in the two seasons.}
\label{tab:table2}
\begin{tabular}{c c c c c}
\hline
February & March & April & November & December \\ 
\hline
4 & \textbf{9} & 6 & 8 &	2 \\
\textbf{8} &	\textbf{11} &	\textbf{8} &	10 &	6 \\
10 &	13 &	12 &	\textbf{12} &	\textbf{12} \\
\textbf{14} &	15 &	14 &	\textbf{20} &	14 \\
     & 17 &	16 &	22 &	16 \\
     & 	19 &	18 &	24 &	\textbf{18} \\
     & 	21 &	20 &	26 &	22 \\
     & 	23 &	\textbf{24} &	28 &	24 \\
     & 	25 &      &      & 	26 \\
     & 	27 &      & 	     & 28 \\
     & 	29 &	     &      & \\ \hline
\end{tabular}
\end{center}
\end{table}

The vector lines were drawn on the imageries, for selected OCM scenes, where the SSC was at its maximum surface extent in the Gulf region stretching towards the open sea. The SSC extent line was restricted to the political border close to the mouth of Indus River delta. The SSC were computed at the northern and southern coastlines of the Gulf (Fig.~\ref{fig:figure1}) and were interpreted using the SSC images (Table~\ref{tab:table3}).

\begin{figure}[htbp]
\centering \includegraphics[height=10cm]{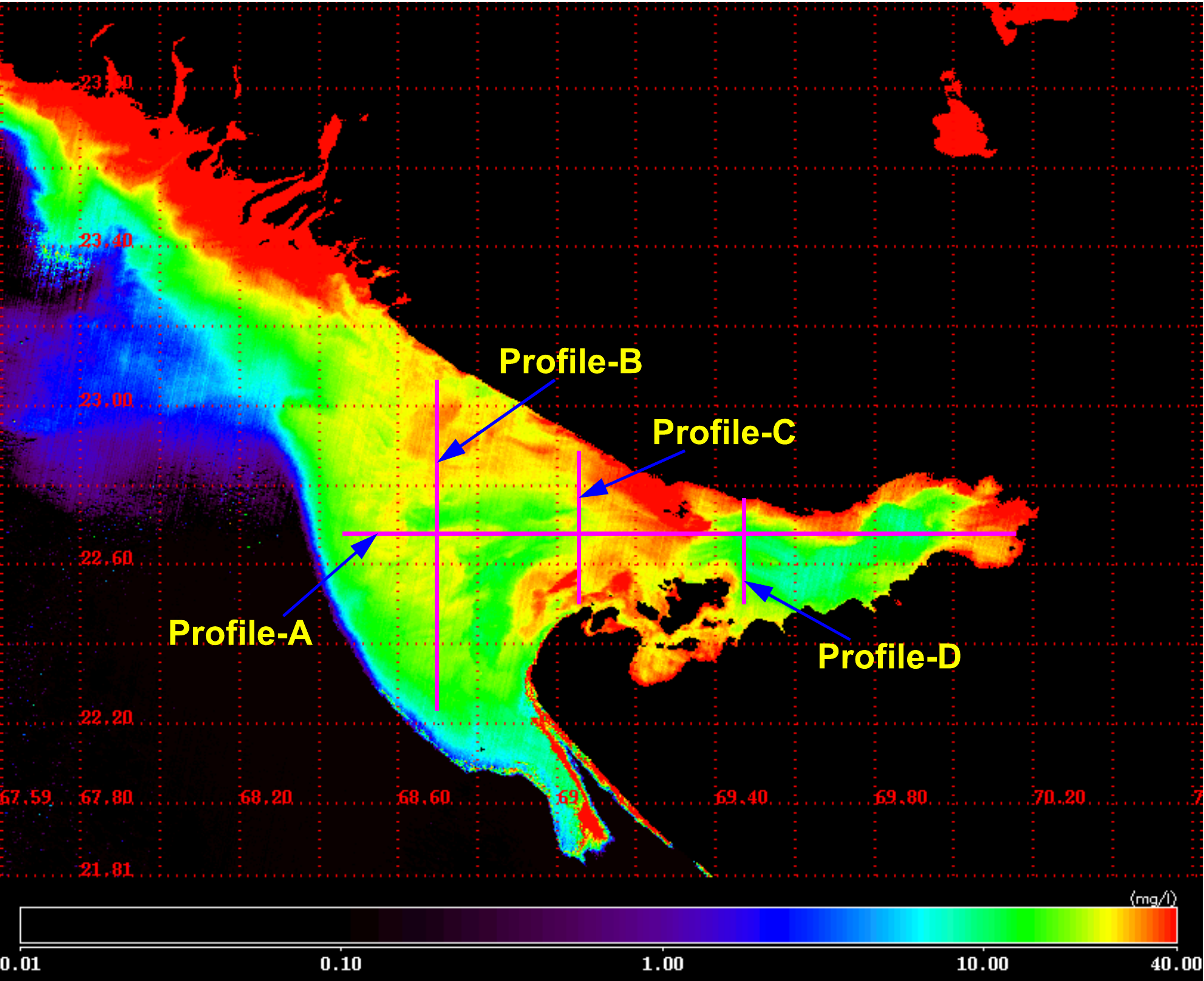} 
\caption{\small Geographic locations of four profiles are shown over an OCM-derived suspended sediment concentration image. Latitude and longitude information of each profile is provided in Table~\ref{tab:table4}.} \label{fig:figure2}
\end{figure}

\begin{table}[h!]
\begin{center}
\caption{OCM-derived suspended sediment concentrations in the northern and southern coasts of Gulf of Kachchh during tidal cycles.}
\label{tab:table3}
\begin{tabular}{l l l}
\hline
Date of OCM pass & Northern coast & Southern coast \\
& (mg/l) & (mg/l ) \\
\hline
\textbf{Pre-monsoon} &  &  \\
08 February 2004 & $\ge$ 5.0 & 10.0--40.0\\
14 February & $\ge$ 20.0 & 5.0--40.0 \\
09 March & $\ge$ 20.0 & $\ge$ 10.0 \\
11 March & $\ge$ 10.0 & $\ge$ 10.0 \\
08 April & $\ge$ 40.0 & $\ge$ 40.0 \\
24 April & $\ge$ 30.0 & 30.0--40.0 \\
\textbf{Post-monsoon} \\
12 November & 20.0--40.0 & $\ge$ 40.0 \\
20 November &	 $\ge$ 20.0 & 20.0--40.0 \\
12 December & 10.0--40.0 & $\ge$ 20.0 \\
18 December & 8.0--35.0	& $\ge$ 8.0 \\
\hline
\end{tabular}
\end{center}
\end{table}
\indent Four profiles were drawn on the OCM-derived SSC images, one parallel to the Gulf mouth and three across the Gulf (Fig.~\ref{fig:figure2}). Geographic location of the profiles is provided in the Table~\ref{tab:table1}. Profile A starts from the west (0 km) extending about 180 km towards the east. Profiles B, C, and D start from the north (0 km) extending to the south ($\sim$80 km). The SSC values were plotted along the profiles and the SSC for different periods (March, April, November, and December) were studied in conjunction with the tidal heights matching with the geo-corrected OCM scenes. To investigate the statistically significant difference between pre- and post-monsoon SSC data, paired samples t-test was done and a p$<$0.008 was found at 95\% confidence interval. No major seismic activity was reported in the region during pre- and post-monsoon in 2004 that could alter the SSC in the coastal waters. Previously, the SSC were observed to increase after the famous Gujarat earthquake of 2001 \cite{singhetal2001}. \\

\begin{table}[h!]
\begin{center}
\caption{Geographic locations of the profiles (shown in Fig.~\ref{fig:figure2}).}
\label{tab:table4}
\begin{tabular}{l l l}
\hline
Profile & Latitude & Longitude \\
\hline
A & 22$^{\circ}$40$^{\prime}$56.57$^{\prime \prime}$N & 68$^{\circ}$17$^{\prime}$30.99$^{\prime \prime}$E--70$^{\circ}$13$^{\prime}$06.56$^{\prime \prime}$E \\
B & 23$^{\circ}$00$^{\prime}$23.08$^{\prime \prime}$N -- 22$^{\circ}$21$^{\prime}$39.83$^{\prime \prime}$N & 68$^{\circ}$37$^{\prime}$32.86$^{\prime \prime}$E \\
C & 22$^{\circ}$54$^{\prime}$41.58$^{\prime \prime}$N -- 22$^{\circ}$29$^{\prime}$18.56$^{\prime \prime}$N & 69$^{\circ}$04$^{\prime}$01.93$^{\prime \prime}$E \\
D & 22$^{\circ}$46$^{\prime}$48.53$^{\prime \prime}$N -- 22$^{\circ}$30$^{\prime}$01.89$^{\prime \prime}$N & 69$^{\circ}$46$^{\prime}$42.61$^{\prime \prime}$E \\
\hline
\end{tabular}
\end{center}
\end{table}

\section*{\normalsize RESULTS}
\subsection*{\normalsize Analysis of SSC Images}
	The satellite remote sensing has made it easier to estimate the surface extent of suspended sediments, which is otherwise very difficult to estimate during the two seasons \cite{gupta2015}. The monsoon usually arrives in India during June and spreads across the country by late June to early July. In the North-East of the Gulf (narrower end of the Gulf, number 5 in Fig.~\ref{fig:figure1}), the SSC is generally very high in the pre-monsoon (Table~\ref{tab:table3}); and in the post-monsoon the SSC remains $>$40.0 mg/l. Being one of the shallowest regions of the Gulf, it experiences a significant influence of tidal forces. In the region between Kalubhar--Mundra (number 8--3) and Jodiya--Kandla (number 6--4), the SSC in pre-monsoon varies between 8.0--35.0 mg/l while during post-monsoon it varies between 10.0--40.0 mg/l. The SSC in pre-monsoon varies between 10.0--40.0 mg/l in the region between Okha--Mandvi (number 9--2) and Kalubhar--Mundra (number 8--3). Longshore movement of suspended sediments is observed near the Okha port (number 9) where, in the post-monsoon, the SSC value varies between the 15.0--40.0 mg/l. As the width of the Gulf increases towards its mouth, the range of variability of SSC increases to 1.0--40.0 mg/l in the west of Okha--Mandvi (number 9--2). This variability remains much higher in the region during both pre- and post-monsoon. In the northern shore (along ports 1--4) of Gulf of Kachchh, the range of variability of SSC (5.0--40.0 mg/l) during pre-monsoon and post-monsoon (10.0--40.0 mg/l) is approximately the same. This is likely due to the re-suspension of sediments along the shore, which is relatively much shallower than the mid-Gulf region. The SSC varies between 5.0--40.0 mg/l for pre-monsoon in the southern shore (along numbers 5--9) and between 20.0--40.0 mg/l during the post-monsoon.

\begin{figure}[htbp]
\centering \includegraphics[height=15cm]{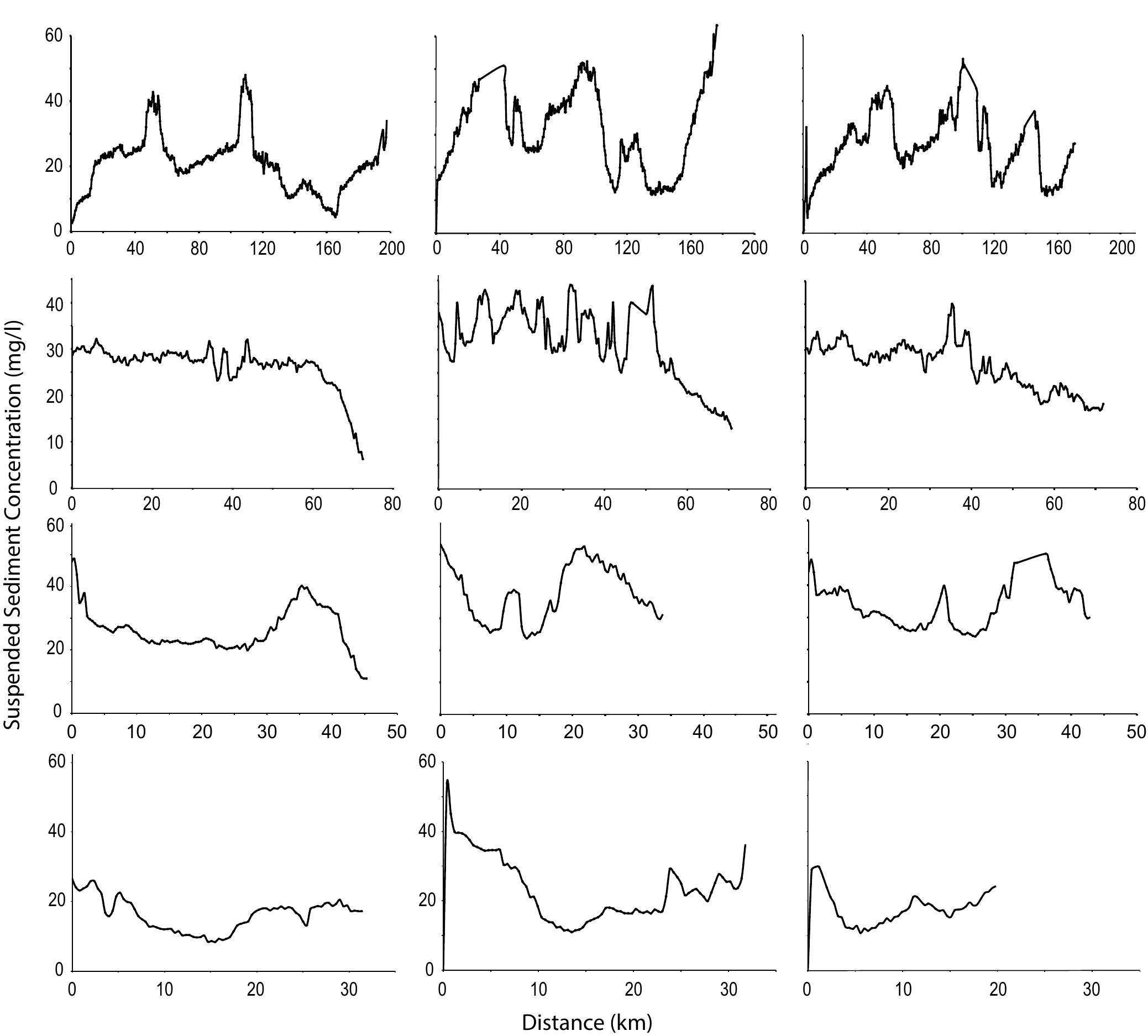} 
\caption{\small Pre-monsoon suspended sediment concentration (SSC) profiles A (0 km at the mouth to $\sim$180 km in the interior of the Gulf), B, C, and D (in consecutive rows from top) plotted using OCM-derived SSC images of 11 March 2004 (left column) and 08 April (mid), and 24 April (right). Profiles B, C, and D start from the north (0 km) extending to the south ($\sim$80 km).}\label{fig:figure3}
\end{figure}

\begin{table}[h!]
\renewcommand{\thetable}{\arabic{table}(a)}
\begin{center}
\caption{Suspended sediment concentration (SSC) along various profiles during pre-monsoon as per tidal cycles.  Statistical paired sample \textit{t}-test between pre- and post-monsoon data reveals p$<$0.008 indicating statistically significant difference.}
\label{tab:table5a}
\begin{tabular}{l l l l l l l}
\hline
\bf Pre-Monsoon & \bf March 2004 \\
Tide condition- Date & High & & & Low- March 11 \\
SSC value (mg/l) & & & & Max. & Min. & Average \\ \hline
Profiles \\					
A & & & & $>$40.0 & 3.0 & 20.7 \\
B & No & & & 35.0 & 6.0 & 26.1 \\
C & Data & & & $>$40.0 & 11.0 & 26.3 \\
D & & & & 27.0 & 8.0 & 16.2 \\ \hline				
 & \bf April 2004 \\
Tide condition- Date & High- April 08 & & & Low- April 24 \\
SSC value (mg/l) & Max. & Min. & Average & Max. & Min. & Average \\ \hline
Profiles \\					
A & $>$40.0 & 12.0 & 30.8 & $>$40.0 & 5.0 & 25.9 \\
B & $>$40.0 & 13.0 & 31.1 & 40.0 & 17.0 & 26.5 \\
C & $>$40.0 & 23.0 & 37.9 & $>$40.0 & 25.0 & 33.3 \\
D & $>$40.0 & 11.0 & 23.2 & 30.0 & 11.0 & 17.4 \\
\hline
\end{tabular}
\end{center}
\end{table}

\begin{table}[h!]
\addtocounter{table}{-1}
  \renewcommand{\thetable}{\arabic{table}(b)}
\begin{center}
\caption{Suspended sediment concentration (SSC) along various profiles during post-monsoon as per tidal cycles. Statistical paired sample \textit{t}-test between pre- and post-monsoon data reveals p$<$0.008 indicating statistically significant difference.}
\label{tab:table5b}
\begin{tabular}{l l l l l l l}
\hline
\bf Post-Monsoon & \bf November 2004 \\
Tide condition- Date & High- November 12 & & & Low- November 20 \\
SSC value (mg/l) & Max. & Min. & Average & Max. & Min. & Average \\ \hline
Profiles \\						
A & 34.0 & 4.0 & 21.6 & $>$40.0 & 2.0 & 22.9 \\
B & 27.0 & 14.0	 & 18.7 & 38.0 & 14.0 & 27.2 \\
C & 33.0 & 11.0	 & 24.4 & $>$40.0 & 12.0 & 27.1 \\
D & 33.0 & 9.0 & 18.3 & $>$40.0 & 2.0 & 18.8 \\
\hline						
 & \bf December  2004 \\
Tide condition- Date & High- December 12 & & & Low- December 18 \\
SSC value (mg/l) & Max. & Min. & Average & Max. & Min. & Average \\ \hline
Profiles \\	
A & 37.0 & 4.0 & 18.2 & 39.0 & 6.0 & 20.1 \\
B & 20.0 & 2.0 & 13.9 & 27.0 & 11.0 & 22.3 \\
C & 34.0 & 12.0 & 23.6 & 28.0 & 17.0 & 22.2 \\
D & 32.0 & 8.0 & 18.6 & 27.0 & 5.0 & 13.6 \\
\hline
\end{tabular}
\end{center}
\end{table}

\subsection*{\normalsize Analysis of Pre-monsoon SSC Profiles}
	Horizontal Profile-A (Fig.~\ref{fig:figure2}, Table~\ref{tab:table4}) is about 180 km long, stretching from the mouth to the interior of the Gulf, which shows a large variability in SSC along the profile (Fig.~\ref{fig:figure3}). The high and low values of SSC along Profile-A on 11 March 2004 are $>$40.0 and 3.0 mg/l, respectively (Fig.~\ref{fig:figure4}). Two regions of highest SSC occur at about 50 km (near Okha) and 110 km (near Kalubhar) from the mouth towards the interior of the Gulf. Similar variability in SSC is observed on 08 April with a maximum at $>$40.0 mg/l and minimum at 12.0 mg/l; and on 24 April with a high at $>$40.0 mg/l and low at 5.0 mg/l (Fig.~\ref{fig:figure3}, Table~\ref{tab:table5a}).

\begin{figure}[htbp]
\centering \includegraphics[height=18cm]{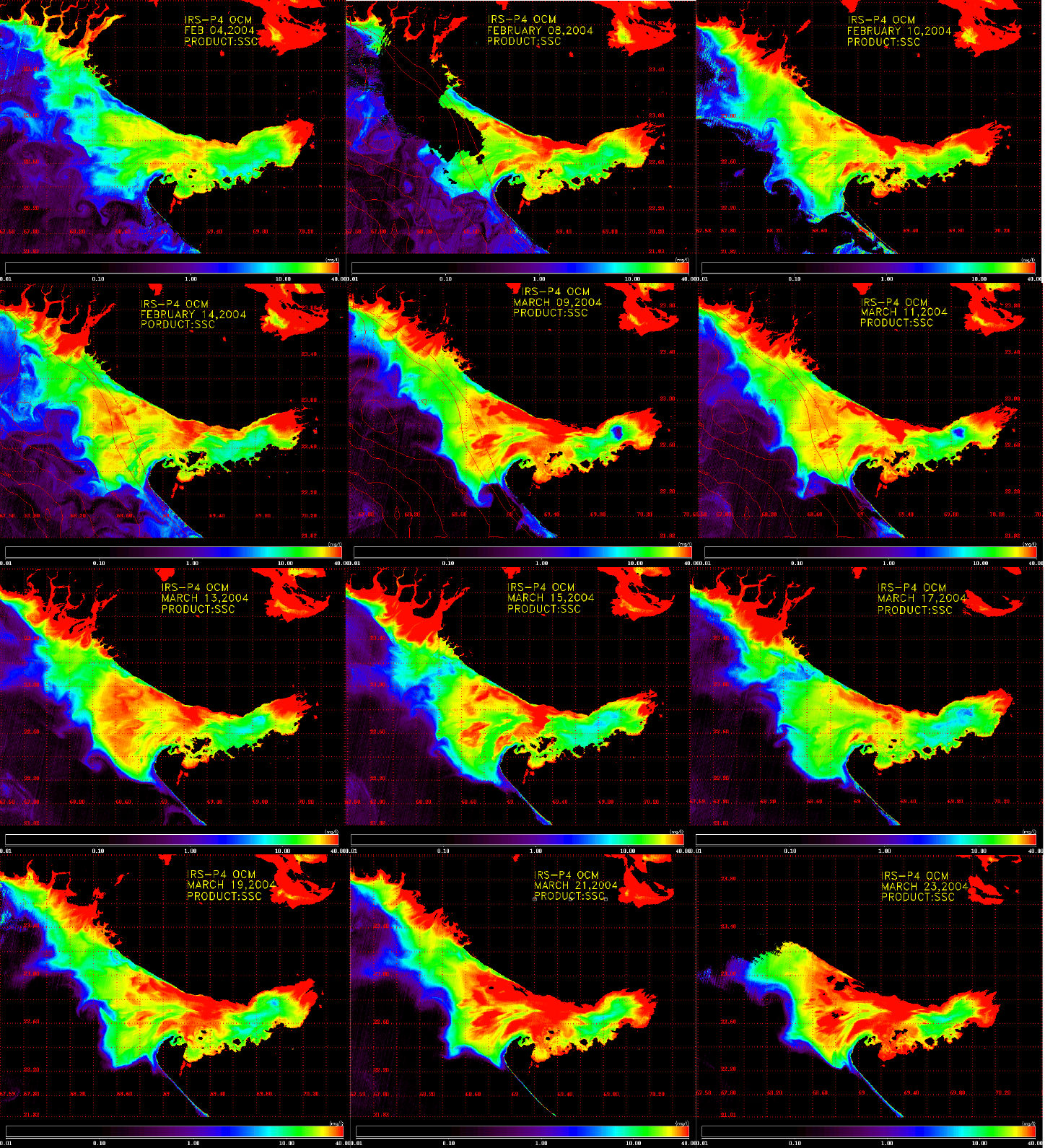} 
\caption{\small Suspended sediment concentration (SSC) images derived using OCEANSAT-1 OCM data of 08 and 14 February, 09 and 11 March, and 08 and 24 April 2004.}\label{fig:figure4}
\end{figure}

\begin{figure}
\renewcommand{\thefigure}{\arabic{figure} (Continued)}
\addtocounter{figure}{-1}
\centering \includegraphics[height=18cm]{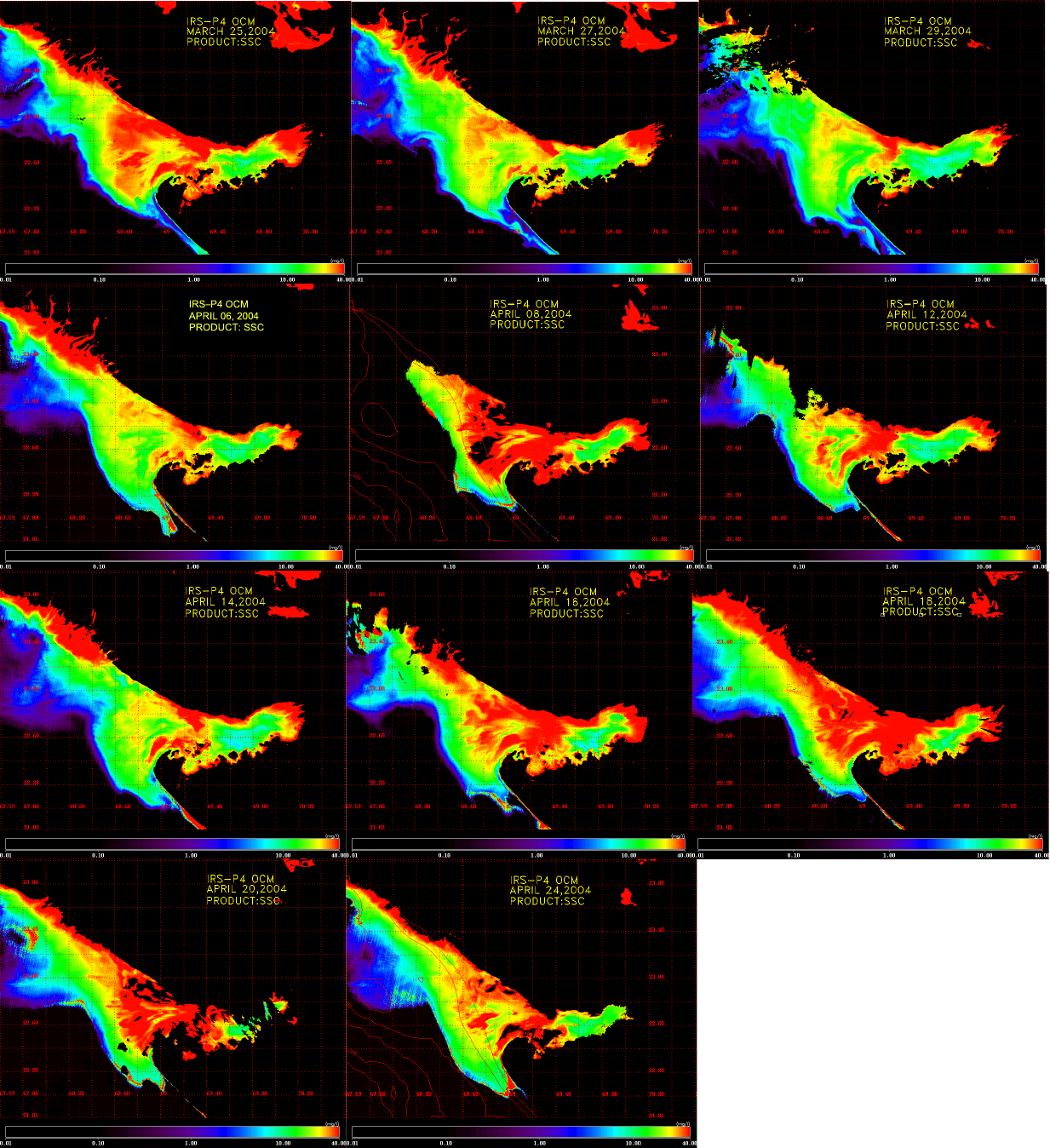} 
\caption{\small Suspended sediment concentration (SSC) images derived using OCEANSAT-1 OCM data of 08 and 14 February, 09 and 11 March, and 08 and 24 April 2004.}\label{fig:figure4cont}
\end{figure}

A vertical Profile-B is taken in the open sea off the mouth of the Gulf about 70 km in length extending from the north to south. A large variability is seen on 11 March 2004 with the SSC maximum at 35.0 mg/l and minimum at 6.0 mg/l. Similar SSC fluctuations were also observed on 08 and 24 April where the SSC varied between $>$40.0--13.0 mg/l which commensurate with the SSC observed on 11 March. These are graphically shown in Fig.~\ref{fig:figure3} and in tabular form in Table~\ref{tab:table5a}. 

\begin{figure}[htbp]
\centering \includegraphics[height=19cm]{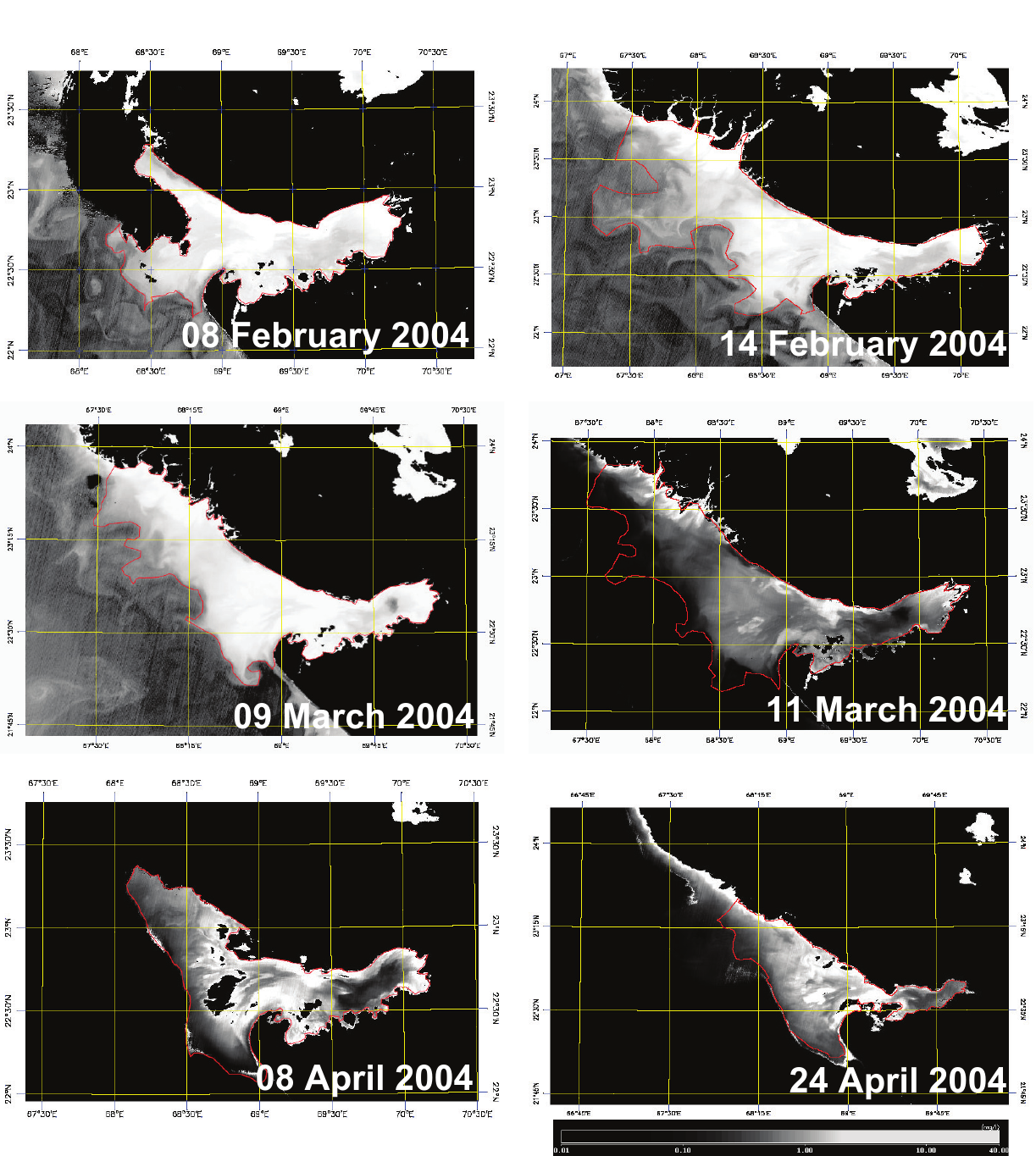} 
\caption{\small The surface extent of suspended sediments (in red) traced using OCM-derived suspended sediment concentration images during pre-monsoon.}\label{fig:figure5}
\end{figure}

Profile-C is about 45 km long stretching vertically from north to south. This profile is close to shallow water areas off Okha port. This is evident through very high SSC range on 08 April ($>$40.0--23.0 mg/l) and 24 April ($>$40.0--25.0 mg/l) (Fig.~\ref{fig:figure3}, Table~\ref{tab:table5a}). The vertical Profile-D lies in the deeper ($\sim$30 m) area inside of the Gulf, which relatively has less SSC. The range of SSC observed along this profile from March to April is 30.0--8.0 mg/l (Fig.~\ref{fig:figure3}, Table~\ref{tab:table5a}). The surface extent of suspended sediments during pre-monsoon is drawn as a vector layer (in red) on the OCM-derived SSC images (Fig.~\ref{fig:figure4} and ~\ref{fig:figure5}). The highest surface extents of suspended sediments during pre-monsoon low and high tides were 21,206.1 and 10,454.7 km$^2$, respectively.

\subsection*{\normalsize Analysis of Post-monsoon SSC Profiles}
	Similar analysis of SSC images was performed for the post-monsoon months, i.e. November and December along the same Profiles A, B, C, and D. The SSC varies between 4.0--34.0 mg/l (12 November), 2.0--$>$40.0 mg/l (20 November), 37.0--4.0 mg/l (12 December), and 6.0--39.0 mg/l (18 December) (Table~\ref{tab:table5b}). It is observed from the comparison of SSC during pre- and post-monsoon along Profile-A that the SSC is less during post-monsoon months.
	
\begin{figure}[htbp]
\centering \includegraphics[height=16cm]{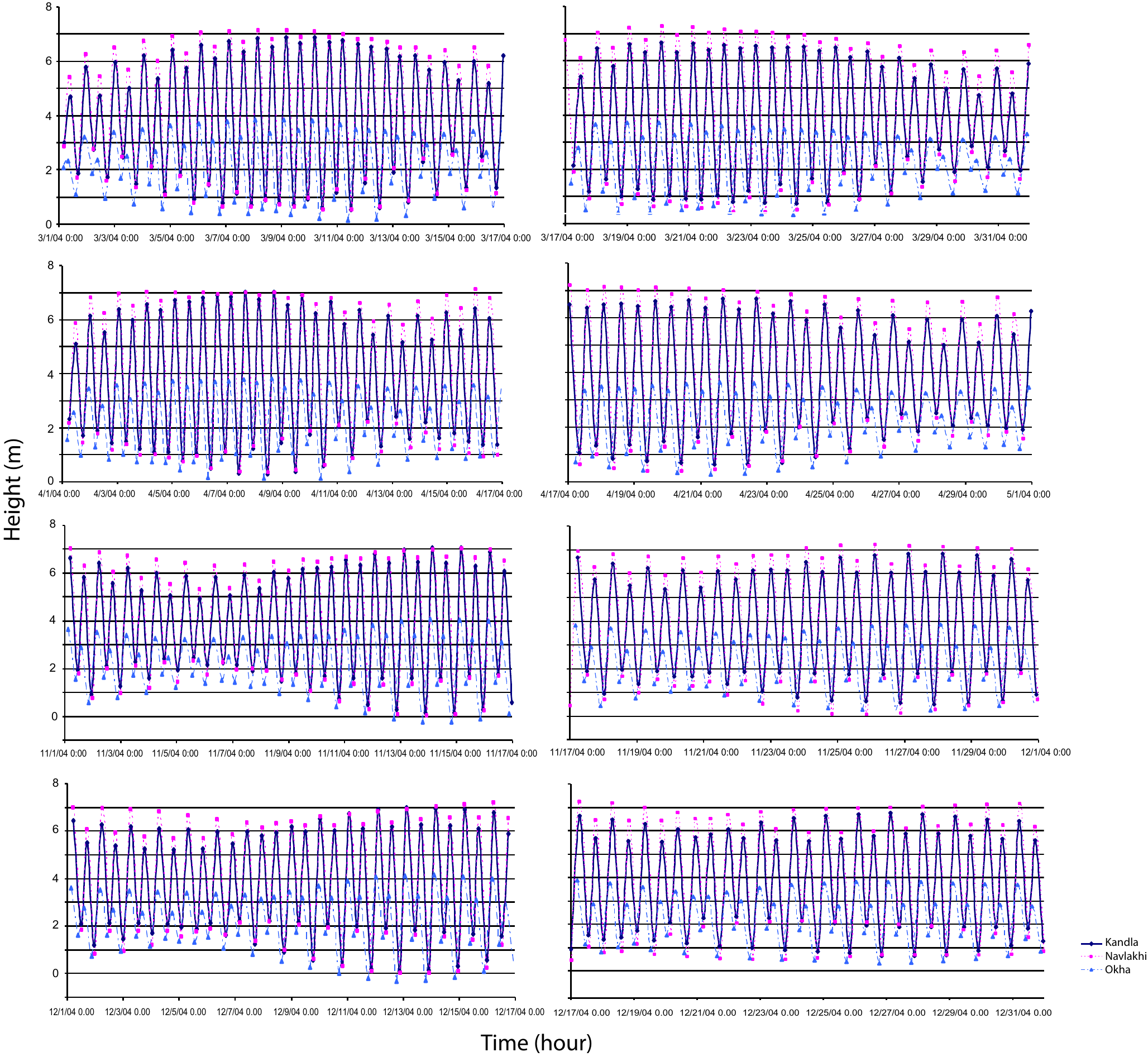} 
\caption{\small Tidal heights at the locations Kandla (number 4 in Fig.~\ref{fig:figure1}), Navlakhi (number 5), and Okha (number 9) during March, April, November, and December 2004.}\label{fig:figure6}
\end{figure}

Along the Profile-B, the SSC shows a range of 14.0--27.0 mg/l (12 November) and 38.0--14.0 mg/l (20 November). Overall SSC on 20 November are higher than that on 12 November because of low tide conditions on 20 November (Fig.~\ref{fig:figure6}--~\ref{fig:figure8}, Table~\ref{tab:table5a}). Profile-C shows a minimum and maximum SSC 11.0 mg/l and $>$40.0 mg/l during November and December much less than that observed during pre-monsoon. The range of SSC along Profile-D during November and December is $>$40.0 and 5.0 mg/l. To sum up, all the profiles for both seasons, it is observed that the SSC during pre-monsoon is significantly higher than that during post-monsoon (Fig.~\ref{fig:figure7} and ~\ref{fig:figure8}). 

\begin{figure}[htbp]
\centering \includegraphics[height=16cm]{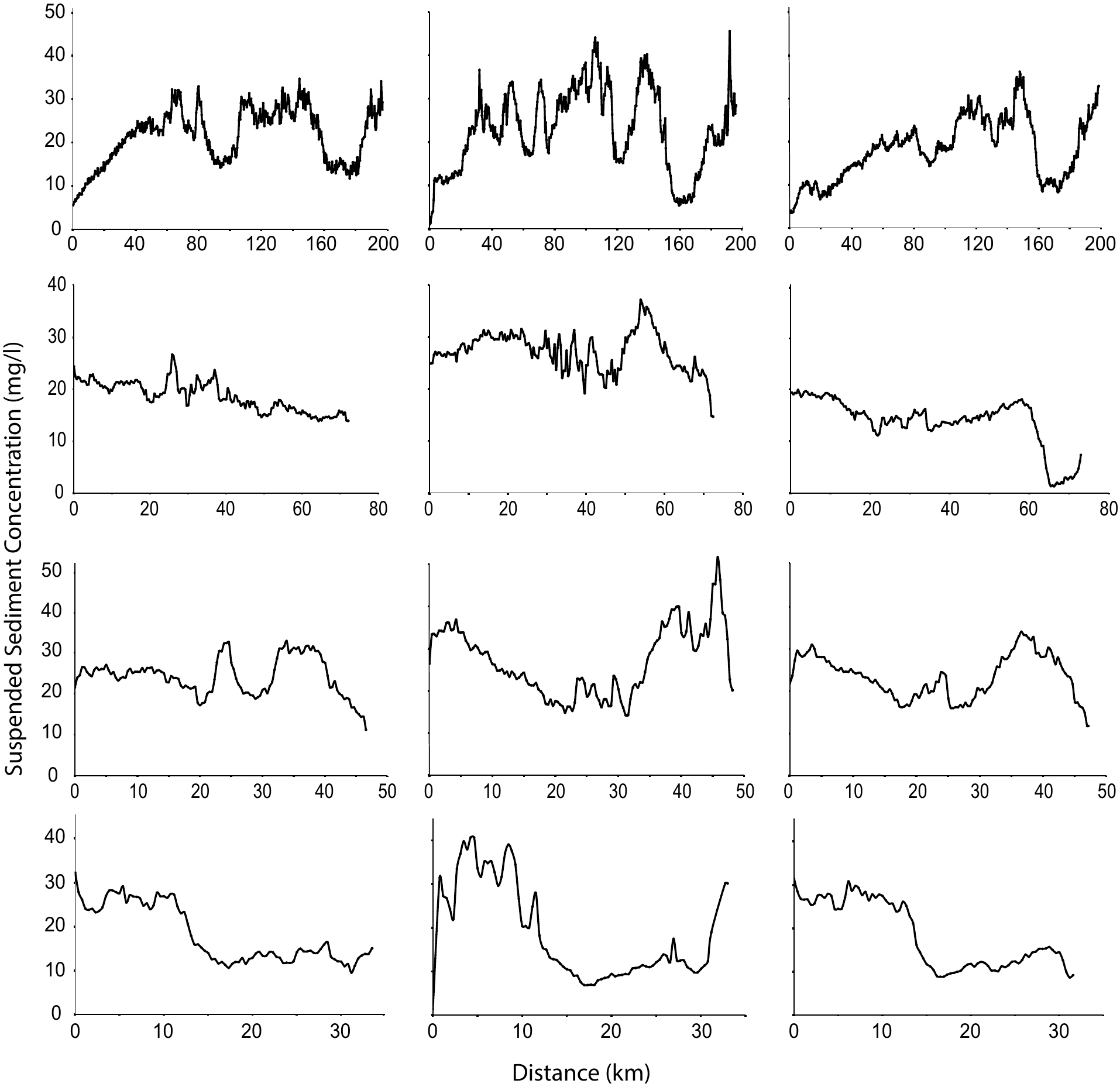} 
\caption{\small Post-monsoon suspended sediment concentration (SSC) Profiles A (0 km at the mouth to $\sim$180 km in the interior of the Gulf), B, C, and D (in consecutive rows from top) plotted using OCM-derived SSC images of 12 November 2004 (left column) and 20 November (mid), and 12 December (right). Profiles B, C, and D start from the north (0 km) extending to the south ($\sim$80 km).}\label{fig:figure7}
\end{figure}

\begin{figure}[htbp]
\centering \includegraphics[height=18cm]{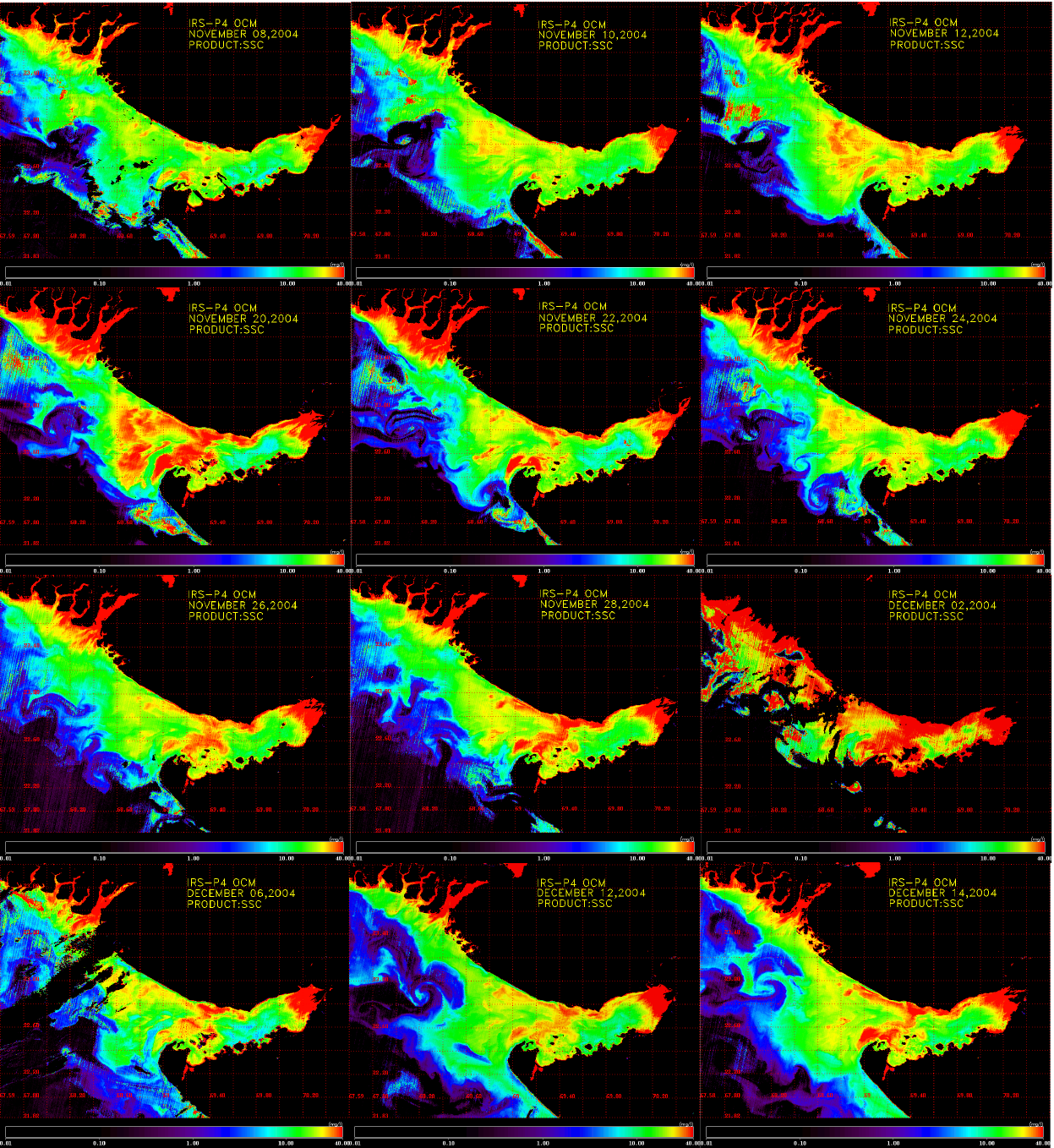} 
\caption{\small Suspended sediment concentration (SSC) images derived using OCEANSAT-1 OCM data of November and December 2004.}\label{fig:figure8}
\end{figure}

\begin{figure}[htbp]
\renewcommand{\thefigure}{\arabic{figure} (Continued)}
\addtocounter{figure}{-1}
\centering \includegraphics[height=9cm]{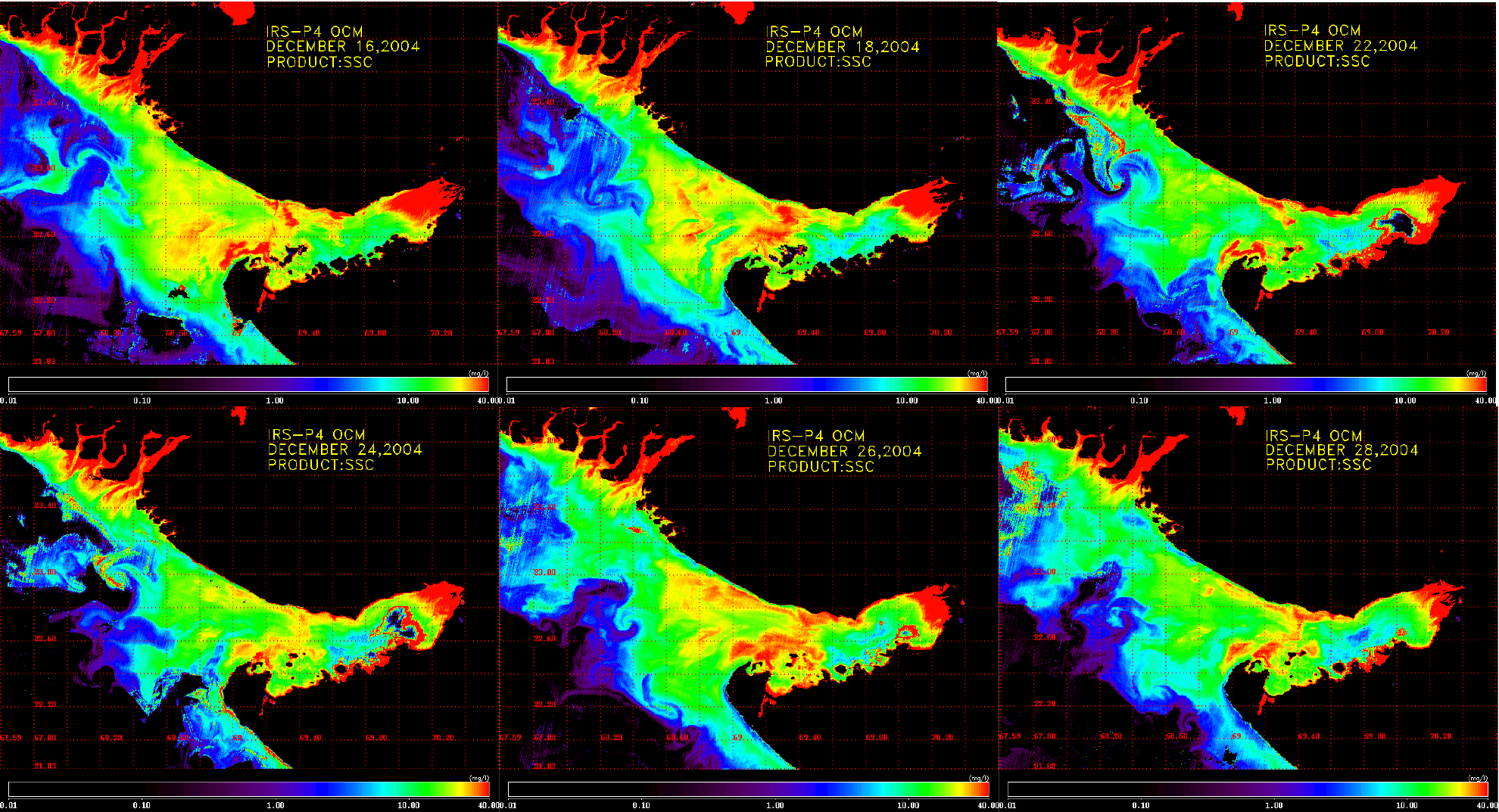} 
\caption{\small Suspended sediment concentration (SSC) images derived using OCEANSAT-1 OCM data of November and December 2004.}\label{fig:figure8cont}
\end{figure}

\begin{table}[h!]
\begin{center}
\caption{The area of surface extent of suspended sediments during pre- and post-monsoon as per tidal cycles.}
\label{tab:table6}
\begin{tabular}{l l l}
\hline
Date & Tide condition & Area of extent of \\
 & & suspended sediments (km$^2$) \\ \hline
\bf Pre-monsoon & & \\
08 February 2004 & High + 15 days & 9,470.3$^a$ \\
14 February & Low + 2 days & 21,206.1 \\
09 March & High -- 1 day & 18,374.5 \\
11 March & Low & 17,996.0 \\
08 April & High & 10,454.7 \\
24 April & Low + 15 days & 10,420.0 \\ \hline
\bf Post-monsoon & & \\
12 November & High	 & 17,165.7 \\
20 November & Low -- 1 day & 13,201.0 \\
12 December & High	 & 18,891.5 \\
18 December & Low -- 3 days & 16,543.6 \\
\hline
\footnotesize $^a$Less area is due to the presence of clouds in the satellite image.
\end{tabular}
\end{center}
\end{table}

\section*{\normalsize DISCUSSION}
\subsection*{\normalsize Interpretation of SSC Images}
	The monsoon in the year 2004 was deficient as per the IMD \cite{rajeevanetal2007} \cite{duttaetal2012}. During the pre-monsoon (February--April), the SSC is observed to be high in all pre-monsoon images (Fig.~\ref{fig:figure4}). The highest SSC is seen on 09 March 2004, which is one day before the occurrence of high tide condition at Okha followed by the low tide occurrence on 11 March. The suspended sediments extend into the open sea (beyond 50 m depth contour) during the low tide condition, which occurred on 14 February and 11 March. The algorithm overestimated the SSC on 08 April due to the presence of clouds in the satellite imagery. The post-monsoon images show low SSC in all the images except on 20 November, which is one day before the low tide occurrence.\\
\indent All tide observations interpreted on a particular day closely match 12:00 noon which is the time of OCEANSAT-1 pass. Fig.~\ref{fig:figure4} shows the increasing tidal height conditions from 04--08 February, and then it decreases till 14 February. Low concentration of suspended sediments is seen (high tide) on 04 February while high SSC is seen on 14 February (low tide).
During 17 and 29 March 2004, very low SSC is observed in the inner Gulf. The tidal heights on these dates were 2.8 m (high tide) and 1.1 m (low tide), respectively. 11 March and 09 March correspond to low and high tides respectively which are also observed in the corresponding high and low SSC images (Fig.~\ref{fig:figure4}). Low SSC is observed on 06 April image with a tidal height of 3.7 m (high tide) showing the increased concentration of suspended sediments. Thus, we observed high and low SSC corresponding to high and low tide conditions during February, March, and April.\\
	\indent Similarly, during the post-monsoon, all SSC images during November and December are in agreement with the corresponding tidal conditions. The SSC images show less concentration of suspended sediments on 12 November and high concentration on 20 November, which correspond to 3.4 m (high tide) and 1.2 m (low tide) tidal heights respectively (Fig.~\ref{fig:figure8}). Similarly, 12 and 28 December SSC images show low concentration of suspended sediments, which correspond to 3.3 and 2.9 m (both high tides), respectively (Fig.~\ref{fig:figure8}). The statistical t-test between pre- and post-monsoon SSC data shows p$<$0.008, which indicates that the two datasets differ statistically significantly. This confirms that the SSC is different in two seasons. It is, however, a topic of interest to investigate which processes have more effect in causing this difference, monsoonal or tidal/surface winds. 

\begin{figure}[htbp]
\centering \includegraphics[height=12cm]{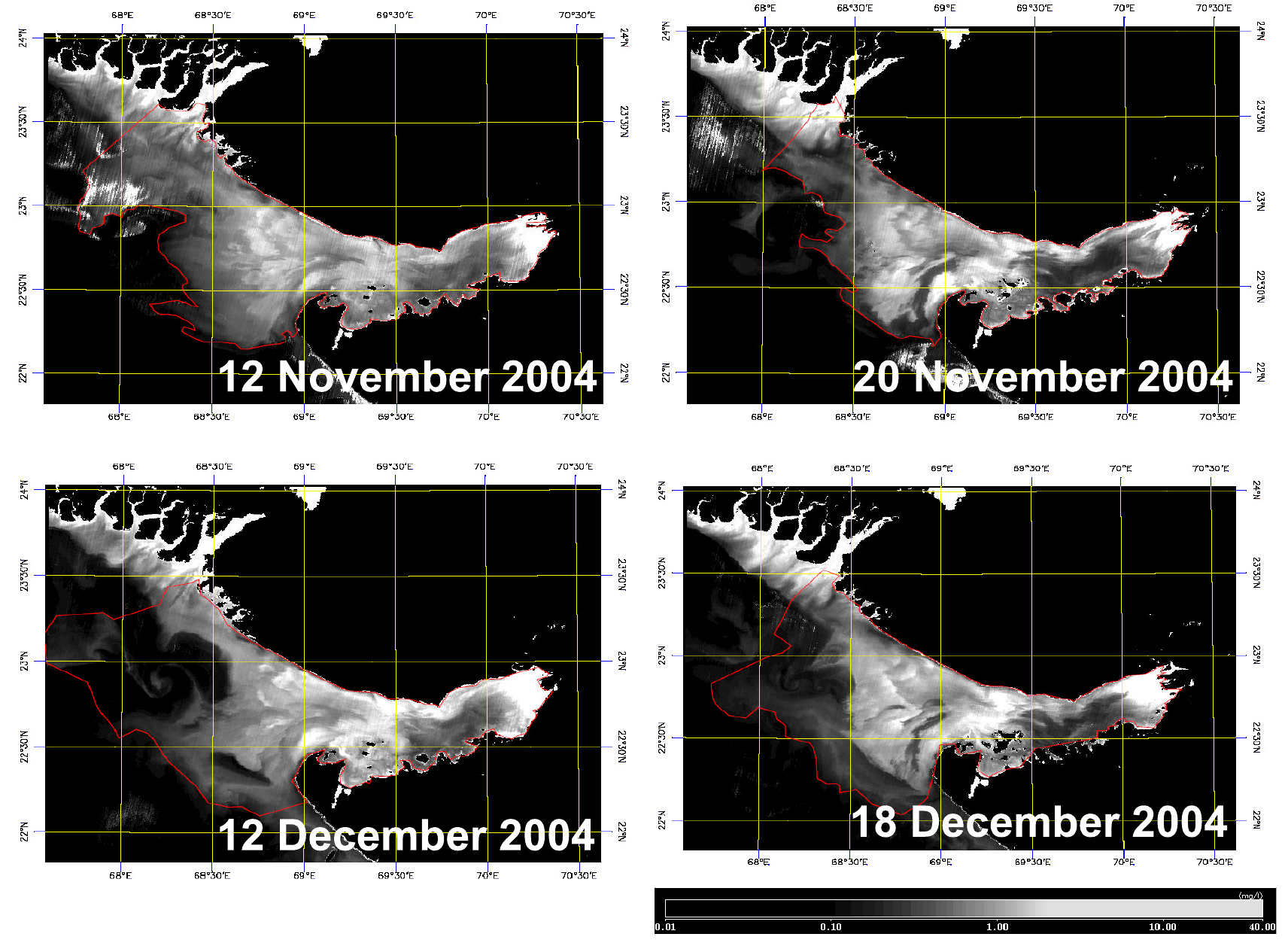} 
\caption{\small The surface extent of suspended sediments (in red) traced using OCM-derived suspended sediment concentration images during post-monsoon.}\label{fig:figure9}
\end{figure}

\indent The surface extent of suspended sediments during post-monsoon is drawn as a vector layer (in red) on the OCM-derived SSC images (Fig.~\ref{fig:figure9}). The highest surface extents of suspended sediments during a post-monsoon low and high tide were 13,201.0 and 18,891.5 km$^2$, respectively. During pre-monsoon, the area of suspended sediments extent is higher in low tide conditions than that during high tide conditions. Contrary to this, the area of extent of suspended sediments during post-monsoon is lower in low tide conditions than that in high tide conditions (Table~\ref{tab:table6}). This is attributed to the increased sediment flux of Indus River, which increases during pre-monsoon months due to the onset of spring in Himalayas (February--April), the origin of Indus River. The annual sediment drainage into the ocean by Indus River is estimated to be much greater than 100$\times$10$^6$ t yr$^{-1}$ \cite{millimanandmeade1983}. Only occasionally does the monsoon extend sufficiently north to cause summer precipitation \cite{garzantietal2005}.  During summer, when snow melts, water discharge increases 20-50 times and sediment load 500-1000 times \cite{ferguson1984}. Glaciofluvial streams attain exceptionally high erosion rates during summer months not so much because of the monsoon rains but because of the rapid melting of winter snow and glacial ice \cite{rehmanetal1997}. The sediment budget of Indus River, the only source of sediment discharge in the Gulf of Kachchh, experiences a low during winter months (November--December).  This is in addition to the increase in SSC due to re-suspension caused by wind and tidal currents in the Gulf. An alternate explanation of the abrupt behavior in surface extents of suspended sediment concentrations during low and high tides in post-monsoon indicate the consideration of re-suspension of bottom sediments in the Gulf of Kachchh and prevailing surface wind conditions. Surface winds and re-suspension of sediments due to tidal currents likely cause the inconsistent SSC observed through OCM images. Otherwise, under normal monsoon conditions, the post-monsoon SSC is expected to be greater than SSC in pre-monsoon. This implies that Gulf of Kachchh is more affected by re-suspension of sediments (due to tidal currents and surface winds) than by the fresh sediment flux brought by Indus River. The surface winds in the Gulf are reported to be 2--5 m/s, which can go up to 10 m/s \cite{pradhanetal2004}. Strong or even normal monsoon conditions could have increased the suspended sediment loads in the Gulf by increasing onshore winds. However, in the absence of a strong monsoon in 2004 \cite{rajeevanetal2007}\cite{duttaetal2012}, the low SSCs expected in post-monsoon support the satellite observations.\\
\indent The unique opportunity to study the role of external factors contributing to the suspended sediments into the Gulf arose due to the absence of normal monsoon in 2004. We make use of the scarcity of monsoon of 2004 to find through satellite data that the SSC in the Gulf increases during pre-monsoon and decreases during post-monsoon, contrary to what is expected in a normal monsoon. However, the SSC is expected to remain consistent throughout the year in the absence of normal monsoon.\\
\indent But the observations tell that the SSC has increased during the pre-monsoon and decreased during the post-monsoon. This clearly points toward the contribution of tidal/wind re-suspension of sediments and/or Himalayan snowmelt (via Indus River) during pre-monsoon (March--April), and/or tidal influences that enhance the SSC. Similarly, a decrease in SSC during post-monsoon is observed on satellite images due to less sediment discharge from the Indus River (less snowmelt in November--December).

\section*{\normalsize CONCLUSIONS}
	In this paper, a comprehensive study of satellite-derived SSC during pre- and post-monsoon was conducted with full-month cycles of tidal responses to suspended sediment dynamics in the Gulf of Kachchh. The analysis is based on 10 (pre- and post-monsoon) SSC images used for sediment extent and SSC computations; and 31 SSC images used for sediment dispersal studies. Tidal data was interpreted in conjunction with the OCM-derived SSC for pre-monsoon (February, March, and April) and post-monsoon (November and December) periods. The area of extent of suspended sediments was derived for the same period using the SSC images. The hypothesis that the SSC increases after the monsoon in the Gulf is found incorrect. In fact, the SSC reduces after the monsoon, and also the effect of monsoon on Gulf sediment dynamics appears non-significant.\\
\indent The following conclusions are drawn from the study: towards objective-1, Gulf of Kachchh undergoes tremendous tidal influences in addition to external/other factors such as winds. The pre- and post-monsoon analyses of the data showed that the Gulf seemed to be more affected by the tidal changes than the monsoonal changes in 2004. We utilize the unique opportunity that arose due to the absence of normal monsoon in 2004 to study the role of external factors other than monsoon contributing to the suspended sediments into the Gulf. This led to a finding that the SSC in the Gulf increases during pre-monsoon and decreases during post-monsoon, contrary to what is expected in a normal monsoon. However, the SSC is expected to remain consistent throughout the year in the absence of normal monsoon, pointing toward the re-suspension of sediments due to tidal and wind forcing. The only little monsoonal influence is seen when Indus River discharges sediments during pre-monsoon due to increased sediment flux from its origin, Himalayas in spring (February--April) as compared to less sediment discharge observed during winter (November--December). The average SSC during pre-monsoon were 30.8 mg/l (high tide) and 24.1 mg/l (low tide); and during post-monsoon 19.7 mg/l (high tide) and 21.8 mg/l (low tide). The highest surface extents of suspended sediments during pre-monsoon were 21,206.1 (low tide) and 10,454.7 km$^2$ (high tide); and during post-monsoon were 13,201.0 (low tide) and 18,891.5 km$^2$ (high tide).\\
\indent Towards objective-2, it is demonstrated that sequential OCM images can be helpful in delineating suspended sediment plumes. The behavior of these plumes and sediment extent depends on myriads of external factors. The inconsistent and abrupt SSC values and surface extents, observed from OCM images, result from the combined influence of the deficient monsoon in 2004, prevailing surface winds, and re-suspension of bottom sediments due to tidal currents in the Gulf. In situ measurements of SSC and its surface extents can be a future validation exercise. The pre-monsoon SSC images show overall high suspended sediments whereas post-monsoon SSC images show comparatively low SSC. The entire suspended sediment dynamics occurs within the 50 m depth contour. The re-suspension of sediments due to wind forcing within the Gulf affects the regular behavior of the sediment dispersal patterns observed on the satellite images. It is hoped that similar studies for multiple and consecutive years, with moderate or 'normal' monsoon under similar tidal conditions, will affirm and/or generalize the findings of this paper. The use of enhanced resolution ocean color satellite data ($<$360 m spatial resolution) for deriving higher SSC ($>$40.0 mg/l) and to study the sediment dispersal patterns and dynamics and its validation is suggested as a future avenue of research.

\section*{\normalsize Acknowledgements} Many thanks are due to Dr. Shailesh R. Nayak (now at Ministry of Earth Sciences), Group Director, Marine and Water Resources Group, SAC (ISRO) for his leadership during this project, and for providing institutional support. The processing of OCM images by Mr. Palak L. Patel is gratefully acknowledged.

    \end{ }
    \end{document}